\documentclass[12pt,a4paper]{article}
\usepackage{amsmath,amssymb,amsfonts,amsthm,amsopn,graphicx}
\usepackage{tikz} 
\usetikzlibrary{arrows}
\allowdisplaybreaks[3]
\numberwithin{equation}{section} 
\newtheorem{theorem}{Theorem}[section]

\newtheorem{statement}[theorem]{Statement}

\newcommand{\ds}{\displaystyle}

\renewcommand{\author}[1]{\large\rm #1\\ \bigskip}
\newcommand{\address}[1]{{\normalsize\it #1\\}\bigskip}
\renewcommand{\title}[1]{\bigskip\bigskip\Large\bf #1\bigskip\bigskip\\}

\usepackage[body={15.5cm,22cm}]{geometry}
\newcommand{\At}{\vartheta^{(1)}}
\newcommand{\Bt}{\vartheta^{(2)}}
\newcommand{\Ct}{\vartheta^{(3)}}

\newcounter{app}
\newcounter{sapp}[app]
\def\theapp{\Alph{app}}
\newcommand{\app}[1]{
\refstepcounter{app}{\vspace{7mm}
\noindent\Large\bf Appendix
\theapp.
 \ #1 \par \vspace{5mm}}
\setcounter{equation}{0}
\def\theequation{\Alph{app}.\arabic{equation}}}

\begin{document}
\vglue 2cm

\begin{center}

\title{On difference equations with `$B$'-type solitons on three dimensional lattice.}
\author{Sergey M.~Sergeev.}

\vspace{.5cm}

\address{Department of Theoretical Physics,
         Research School of Physics and Engineering,\\
    Australian National University, Canberra, ACT 0200, Australia\\
    and\\
   Faculty of Science and Technology, \\
   University of Canberra, Bruce ACT 2617, Australia }


\end{center}
\begin{abstract}
In this paper we discuss an example of classical integrable equation with rather unusual `B'-type Kadomtsev-Petviashvili (KP) soliton hierarchy.
\end{abstract}



\section{Zero curvature problem: the Euler angles.}

Let $\boldsymbol{g}$ be an element of group $SO(3)$. It has the unique decomposition into a sequence of simple rotations called the Euler decomposition,
\begin{equation}\label{Euler}
\boldsymbol{g}
\;=\;
\underbrace{\left(\begin{array}{ccc}
\cos\phi & -\sin\phi & 0 \\
\sin\phi & \cos\phi & 0\\
0 & 0 & 1
\end{array}
\right)}_{\ds X_{12}(\phi)}
\underbrace{\left(\begin{array}{ccc}
\cos\theta & 0 & -\sin\theta \\
0 & 1 & 0 \\
\sin\theta & 0 & \cos\theta 
\end{array}
\right)}_{\ds X_{13}(\theta)}
\underbrace{\left(\begin{array}{ccc}
1 & 0 & 0\\
0 & \cos\psi &  -\sin\psi \\
0 & \sin\psi & \cos\psi 
\end{array}
\right)}_{\ds X_{23}(\psi)}
\end{equation}
where $\phi,\theta,\psi$ are the Euler angles.
\\

The same element $\boldsymbol{g}$ can be parametrised by another triple of Euler angles corresponding to the opposite sequence of matrices $X_{\alpha\beta}$,
\begin{equation}\label{ALP}
\boldsymbol{g}\;=\;X_{12}(\phi)X_{13}(\theta)X_{23}(\psi)\;=\;
X_{23}(\psi')X_{13}(\theta')X_{12}(\phi')\;.
\end{equation}
Equation (\ref{ALP}) thus defines uniquely the map
\begin{equation}\label{Rmap}
R\;:\;\; \phi,\theta,\psi\;\to\;\phi',\theta',\psi'\;.
\end{equation}
Since (\ref{ALP}) can be seen as an auxiliary linear problem in $3d$, the map defines a classical lattice integrable model in $3d$. 
  
\bigskip

\noindent{\textbf{Remarks.}} Equation (\ref{ALP}) is a particular case of Korepanov $3d$ linear problem \cite{K1,K2}. Euler angles decomposition as an example of a scalar solution to the Functional Tetrahedron equation was observed in \cite{SF,KKS}. See also \cite{KV,BS1,BMS} for the quantisation of the Korepanov model. Quantum reductions were studied in \cite{S-BKP}. The aim of this paper is to discuss the result of reduction of the classical Korepanov's general case to the particular case of Euler angles.

\section{Lattice equation of motion.}

Let $\vartheta\;:\;\mathbb{Z}^2\;\to\;\mathbb{C}$ be the scalar field variable on cubic lattice. Define the lattice shift notations by
\begin{equation}\label{lvar}
\vartheta\;=\;\vartheta(n_1,n_2,n_3)\;,\quad \vartheta_1\;=\;\vartheta(n_1+1,n_2,n_3)\;,\quad \textrm{etc.}
\end{equation}
Equation (\ref{ALP}) provides the lattice equation of motion by means of identification
\begin{equation}\label{tau2}
\cos^2\psi\;=\;\eta_1^2\, \frac{\vartheta\vartheta_{23}}{\vartheta_2\vartheta_3}\;,\quad
\cos^2\theta'\;=\;\eta_2^2\, \frac{\vartheta\vartheta_{13}}{\vartheta_{1}\vartheta_{3}}\;,
\quad
\cos^2\phi\;=\;\eta_3^2\, \frac{\vartheta\vartheta_{12}}{\vartheta_1\vartheta_2}\;,
\end{equation}
and
\begin{equation}
\cos^2\psi'\;=\;\eta_1^2\, \frac{\vartheta_1\vartheta_{123}}{\vartheta_{12}\vartheta_{13}}\;,\quad
\cos^2\theta\;=\;\eta_2^2\, \frac{\vartheta_2\vartheta_{123}}{\vartheta_{12}\vartheta_{13}}\;, \quad
\cos^2\phi'\;=\;\eta_3^2\, \frac{\vartheta_3\vartheta_{123}}{\vartheta_{13}\vartheta_{23}}\;.
\end{equation}
Here $\eta_k$ are constant coefficients.
The single non-trivial relation in (\ref{ALP}) in new the lattice variables is
\begin{equation}\label{EU}
\eta_1^{2}\eta_2^{2}\eta_3^{2}\,\vartheta\vartheta_{123}^{} \;=\; 
\eta_1^2\,\vartheta_1^{}\vartheta_{23}^{}\,+\,
\eta_2^2\,\vartheta_2^{}\vartheta_{13}^{}\,+\,
\eta_3^2\,\vartheta_3^{}\vartheta_{12}^{}\;-\;
2\;\frac{\vartheta_1^{}\vartheta_2^{}\vartheta_3^{}-H}{\vartheta}\;,
\end{equation}
where
\begin{equation}\label{phi}
H^2\;=\;(\vartheta_1^{}\vartheta_2^{}\,-\,\eta_3^2\vartheta\vartheta_{12}^{})
(\vartheta_1^{}\vartheta_3^{}\,-\,\eta_2^2\vartheta\vartheta_{13}^{})
(\vartheta_2^{}\vartheta_3^{}\,-\,\eta_1^2\vartheta\vartheta_{23}^{})\;.
\end{equation}

\section{`AKP', `BKP' and `KDV' solitons.}

In this section we remind the definition of ``rational $\Theta$-functions''.

Rational ``$\Theta$-function'' of KP type `A' with ``soliton modes''
\begin{equation}
(F,x,y),\quad (F',x',y'),\quad \dots
\end{equation}
is given by
\begin{equation}\label{Two-sol}
\Theta([F,F',\dots]) \;=\; 1\;+\;F\;+\;F'\;+\;F\,F'\,S_A\;+\;\cdots
\end{equation}
where
\begin{equation}\label{A-phase}
S_A\;=\;\frac{(x-x')(y-x')}{(x-y')(y-x')}\;.
\end{equation}
Expression (\ref{Two-sol}) is written for two soliton modes. The general expression is uniquely defined by the principle
\begin{quote}
\emph{For a $g$-soliton $\Theta$-function with soliton amplitudes $F,F',\dots, F^{(g-1)}$,
\begin{equation}
\Theta\;\to\;\Theta'\;=\;\partial \Theta_g/\partial F^{(g-1)}
\end{equation}
is a $(g-1)$-soliton $\Theta$-function with properly re-defined amplitudes.}
\end{quote}

\bigskip
\noindent
In what follows, define 
\begin{equation}\label{E}
E(x,y)\;=\;\frac{x-y}{x+y}\;,
\end{equation}
and the exponents
\begin{equation}\label{exps}
\omega_k^{}\;=\;\frac{E(x,p_k)}{E(y,p_k)}\;,\quad 
\omega_k'\;=\;\frac{E(x',p_k)}{E(y',p_k)}\;,\quad \textrm{etc.}
\end{equation}
These exponents play important role in soliton solutions of difference equations. In particular, it is convenient to define the shifts of amplitudes by
\begin{equation}\label{shifts}
\Theta_k\;=\;\Theta([F\omega_k^{},F'\omega_k',\dots])\;.
\end{equation}
\\
Consider next a $(2g)$-soliton $\Theta$-function with the modes
\begin{equation}
(F,x,y)\;,\;\; (F',x',y')\;,\;\; \dots\;\; 
(F^{(2g-1)},x^{(2g-1)},y^{(2g-1)})\;.
\end{equation}
\\
The key observation is that
\begin{equation}\label{pairing1}
\omega_k(x,y)\;=\;\omega_k(x^{(g)},y^{(g)})\;,\quad 
\omega_k(x',y')\;=\;\omega_k(x^{(g+1)},y^{(g+1)})\;,\quad 
\textrm{etc.}
\end{equation}
if 
\begin{equation}\label{pairing2}
x^{(g)}\;=\;-y\quad \textrm{and}\quad y^{(g)}\;=\;-x\;,\quad 
x^{(g+1)}\;=\;-y'\quad \textrm{and}\quad y^{(g+1)}\;=\;-x'\;,\quad 
\textrm{etc.}
\end{equation}
Let us denote such situation by the collection of pairs
\begin{equation}
(F,\tilde{F},x,y),\quad (F',\tilde{F}',x',y')\;,\quad\dots\;,
\end{equation}
where $\tilde{F}$ stands for $F^{(g)}$, etc.
Three remarkable factorisation happens in this case. The first one is

\begin{statement}
For the pairs of the amplitudes chosen as
\begin{equation}
\Theta\;=\;\Theta([\left(\frac{2x}{(x+y)}f,\frac{2y}{(x+y)}f\right),\left(\frac{2x'}{(x'+y')}f',\frac{2y'}{(x'+y')}f'\right)\dots])\;,
\end{equation}
the $\Theta$-function factorises into the complete square,
\begin{equation}\label{factor1}
\Theta\;=\;\tau^2\;,
\end{equation}
where $\tau$ is the `B'-type theta-function,
\begin{equation}
\tau([f,f',\dots])\;=\;1\;+\;f\;+\;f'\;+\;f\,f'\,S_B\;+\;\cdots\;
\end{equation}
with
\begin{equation}
S_B\;=\;\frac{E(x,x')E(y,y')}{E(x,y')E(y,x')}\;.
\end{equation}
\end{statement}

\bigskip

\noindent
Another example of the factorisation is
\begin{statement}
If the pairs of the amplitudes are chosen as
\begin{equation}\label{scenario2}
\Theta\;=\;\Theta([\left(\frac{2x}{(x+y)}\frac{(y+p_k)}{(x+p_k)} f,
\frac{2y}{(x+y)}\frac{(x-p_k)}{(y-p_k)} f\right),\cdots])\;,
\end{equation}
ane has
\begin{equation}\label{factor2}
\Theta\;=\;
\tau\tau_k\;,
\end{equation}
where
\begin{equation}
\tau_k\;=\;\tau([f\omega_k^{},f'\omega_k',\dots])\;,
\end{equation}
cf. with (\ref{shifts}). 
\end{statement}

\noindent
An example of application of such factorisation will be given in the Appendix A.
\\

\noindent
The third example of complete square factorisation will be given in the next section.

\bigskip

\noindent
In conclusion of this section we would like to mention KDV solitons.
The KDV rational $\Theta$-function correspond to the A-type one with 
\begin{equation}\label{KDV}
y\;=\;-x\;,\quad \textrm{etc., so that} \quad
S_{\textrm{KDV}}\;=\;E(x,x')^2\;.
\end{equation}

\section{Solution to \ref{EU}.}

Let us fix parameters $\eta_k$ in (\ref{EU}) by
\begin{equation}\label{etas}
\eta_1\;=\;E(p_2,p_3)\;,\quad \eta_2\;=\;E(p_3,p_1)\;,\quad \eta_3\;=\;E(p_1,p_2)\;,
\end{equation}
where $E$ is defined by (\ref{E}).

Equation (\ref{EU}) has two types of solution. The first one is the KDV type, see (\ref{KDV}). In the KDV case
\begin{equation}
\omega_k^{}\;=\;\varepsilon_k^2\;,\quad \textrm{where}\quad \varepsilon_k^{}\;=\;E(x,p_k)\;.
\end{equation}
Note that in all cases the lattice shifts (\ref{lvar}) are identified with (\ref{shifts}).

\begin{statement}
Solution to (\ref{EU}) is given by
\begin{equation}
\vartheta\;=\;\Theta_{\textrm{KDV}}\;.
\end{equation}
The factorisation of $H$ is based on the identity
\begin{equation}\label{factor3}
\vartheta_1^{}\vartheta_2^{}\;-\;\eta_3^2\vartheta\vartheta_{12}^{}\;=\;(1-\eta_3^2) A_{12}^2\;,
\end{equation}
where
\begin{equation}
A_{12}^{}\;=\;\Theta_{\textrm{KDV}}([F\varepsilon_1^{}\varepsilon_2^{},F'\varepsilon_1'\varepsilon_2',\dots])\;.
\end{equation}
Thus, the radical $H$ in (\ref{EU}) becomes
\begin{equation}\label{H}
H\;=\;\frac{8p_1p_2p_3}{(p_1+p_2)(p_2+p_3)(p_3+p_1)} A_{12} A_{13} A_{23}\;.
\end{equation}
\end{statement}

\bigskip

\noindent
The other solution to (\ref{EU}) is implicitly the `B'-type one,

\begin{statement}
Solution of (\ref{EU}) is given by
\begin{equation}
\vartheta\;=\;\Theta([(f,f),\dots])
\end{equation}
where we have to use `A'-type notation but with `B'-type pairing content. The factorisation (\ref{factor3}) is valid with
\begin{equation}
A_{12}\;=\;\Theta([\left(\frac{(x-p_1)(y+p_2)}{(y-p_1)(x+p_2)}f,
\frac{(y+p_1)(x-p_2)}{(x+p_1)(y-p_2)}f\right),\dots])\;.
\end{equation}
The radical $H$ in (\ref{EU}) is given by the expression (\ref{H}).
\end{statement}

\app{An example of (\ref{factor2})-factorisation.}

There is another type of an auxiliary linear problem in $3d$. It provides another integrable system, \cite{S0,S1,S2}. In the Lagrangian variables corresponding equations of motion is a set of cubic equations, \cite{S3,S4,S5},
\begin{equation}\label{main}
\left\{
\begin{array}{l}
\ds 
\eta_1^{}\, \At_{\phantom{1}} \Bt_3 \Ct_2 \;+\; \eta_2^{}\, \At_3 \Bt_{\phantom{1}} \Ct_2 \;+\; \eta_3^{}\, \At_2 \Bt_3 \Ct_{\phantom{1}} \;+\; \eta_0^{}\, \At_{23} \Bt_{\phantom{1}} \Ct_{\phantom{1}} \;=\;0\;,\\
[7mm]
\ds 
\eta_1^{}\, \At_{\phantom{1}} \Bt_3 \Ct_1 \;+\; \eta_2^{}\, \At_3 \Bt_{\phantom{1}} \Ct_1 \;+\; \eta_3^{}\, \At_3 \Bt_1 \Ct_{\phantom{1}} \;+\; \eta_0^{}\, \At_{\phantom{1}} \Bt_{13} \Ct_{\phantom{1}} \;=\;0\;,\\
[7mm]
\ds 
\eta_1^{}\, \At_{\phantom{1}} \Bt_1 \Ct_2 \;+\; \eta_2^{}\, \At_2 \Bt_{\phantom{1}} \Ct_1 \;+\; \eta_3^{}\, \At_2 \Bt_1 \Ct_{\phantom{1}} \;+\; \eta_0^{}\, \At_{\phantom{1}} \Bt_{\phantom{1}} \Ct_{12} \;=\;0\;,
\end{array}\right.
\end{equation}
where the $\eta_k$ are given by (\ref{etas}) and 
\begin{equation}
\eta_0\;=\;\eta_1\eta_2\eta_3\;.
\end{equation}

\begin{statement}
This system has the `A'-type solution
\begin{equation}
\vartheta^{(j)}\;=\;\Theta([\frac{(y+p_j)}{(x+p_j)} F,\dots])\;.
\end{equation}
\end{statement}

\noindent
Assuming here the pairing (\ref{pairing2}) with (\ref{scenario2}), one obtains the factorisation (\ref{factor2}), so that each of equations (\ref{main}) reduces to the Hirota-Miwa equation \cite{HM}, the defining identity for the B-type:
\begin{equation}\label{HM}
\eta_1\,\tau_1\tau_{23}\;+\;\eta_2\,\tau_2\tau_{13}\;+\;\eta_3\,\tau_3\tau_{12}\;+\;\eta_0\,\tau\tau_{123}\;=\;0\;.
\end{equation}

\bigskip

\noindent
\textbf{Acknowledgement.} I would like to thank Vladimir Bazhanov and Vladimir Mangazeev for valuable discussions.
The support of the Australian Research Council grant 
DP190103144 is also acknowledged.


\begin{thebibliography}{100}

\bibitem{K1}
I.\ G.\ Korepanov, \emph{``Tetrahedron equation and algebraic geometry''}, Zap. Nauchn. Sem. S.-Peterburg. Otdel. Mat. Inst. Steklov. (POMI) \textbf{209} (1994), no. Voprosy Kvant. Teor. Polya i Statist. Fiz. \textbf{12}, 137-149.

\bibitem{K2}
I.\ G.\ Korpanov, \emph{``Fundamental mathematical structures of integrable models''}. Theoretical and Mathematical Physics \emph{118} (1999) no 3, 405-412.

\bibitem{SF}
S.\ M.\ Sergeev, \emph{``Solutions of the Functional Tetrahedron Equation related to the local Yang-Baxter equation for the ferro-electric condition''}, Lett. Math. Phys. \textbf{45} (1998), 113-119.


\bibitem{KKS}
R.\ M.\ Kashaev, I.\ G.\ Korepanov and S.\ M.\ Sergeev, \emph{``Functional Tetrahedron Equation''}. Theoretical and Mathematical Physics \textbf{117} (1998), 370-384.

\bibitem{KV}
M.\ M.\ Karpanov and V. A. Voevorsky, \emph{``$2$-categories and Zamolodchikov Tetrahedra Equations''}. Algebraic groups and their generalisations: quantum and infinite dimensional methods (University Park, PA, 1991). Proc. Sympos. Pure Math., vol. 56, Amer. Math. Soc., Providence, RI, 1994, pp. 177-259.

\bibitem{BS1}
V.\ V.\ Bazhanov and S.\ M.\ Sergeev, \emph{``Zamolodchikov's tetrahedron equation and hidden structure of quantum groups''}. J. Phys. A \textbf{39} (2006) no. 13, 3295-3310.

\bibitem{BMS}
V.\ V.\ Bazhanov, V.\ V.\ Mangazeev and S.\ M.\ Sergeev, \emph{``Quantum geometry fo $3$-dimensional lattices''}, J. Stat. Mech. (2008), P07006.

\bibitem{S-BKP}
S.\ Sergeev, \emph{``Ground states of the Heisenberg evolution operator in discrete three-dimensional spacetime and quantum discrete BKP equations''}. J. Phys. A: Math. Theor. \textbf{42} (2009) 295207.

\bibitem{S0}
S.\ Sergeev, \emph{``Quantization of three-wave equations''}. J. Phys. A: Math. Theor. \textbf{40} (2007) pp 12709–12724.

\bibitem{S1}
S.\ M.\ Sergeev, \emph{``3D symplectic map''}. Phys. Lett.\textbf{ A 253} (1999) pp 145-150.

\bibitem{S2}
S.\ M.\ Sergeev, \emph{``Quantum $2 + 1$ evolution model''}. J. Phys. A: Math. Gen. \textbf{32} (1999) pp 5693-5714 


\bibitem{S3}
V.\ V.\ Mangazeev and S.\ M.\ Sergeev, \emph{``The continuous limit of the triple $\tau$-function model''}. Theoretical and Mathematical Physics \textbf{129} (2001) pp 317-326.

\bibitem{S4}
S.\ M.\ Sergeev, \emph{``Solitons in a $3d$ integrable model''}.  Phys. Lett. \textbf{A 265} (2000) pp 364-368.

\bibitem{S5}
S.\ M.\ Sergeev, \emph{``On exact solution of a classical 3D integrable model''}. J. Nonlinear Math. Phys. \textbf{1} (2000) pp 57-72. 

\bibitem{HM}
T.\ Miwa, \emph{``On Hirota's difference equation''}. Proc. Japan Acad. Ser. A Math. Sci. \textbf{58} (1982) no. 1, 9-12.




\end{thebibliography}
\end{document}